\documentclass[twocolumn,showpacs]{revtex4}

\usepackage[dvips]{graphicx}
\usepackage{amsmath, amssymb}
\usepackage{dcolumn}
\usepackage{latexsym}

\begin{document}

\preprint{}    

\title{Transition to turbulence in particulate pipe flow}
\author{J.-P. Matas$^{1}$, J. F. Morris$^{2}$, and \'E.
Guazzelli$^{1}$} \affiliation{$^{1}$ IUSTI - CNRS UMR 6595,
Polytech'Marseille, Technop\^ole de Ch\^ateau-Gombert, 13453 Marseille
Cedex 13, France.\\ $^{2}$ School of Chemical Engineering, Georgia
Institute of Technology, Atlanta, Georgia 30332, USA}

\begin{abstract}
We investigate experimentally the influence of suspended particles on
the transition to turbulence.  The particles are monodisperse and
neutrally-buoyant with the liquid.  The role of the particles on the
transition depends both upon the pipe to particle diameter ratios and
the concentration.  For large pipe-to-particle diameter ratios the
transition is delayed while it is lowered for small ratios.  A scaling
is proposed to collapse the departure from the critical Reynolds 
number
for pure fluid as a function of concentration into a single master
curve.
\end{abstract}

\pacs{47.20.-k, 83.80.Hj, 83.50.Ha}

\maketitle

More than a century after Reynolds' work \cite{Reynolds},
understanding how turbulent regions grow in a pipe and bring the
laminar Poiseuille flow to fully developed turbulence is still not
completely achieved.  Above a critical Reynolds number, the 
laminar flow is observed to be unstable, turbulent regions grow and
are convected in the pipe.  This flow regime is called intermittent. 
When the flow rate is further increased, the flow becomes fully
turbulent \cite{Tritton}.  In fact, the transition happens to be
subcritical and the flow is linearly stable for all flow rates
\cite{Drazin}.  A finite amplitude perturbation is needed to trigger
the transition and the critical Reynolds depends upon its amplitude. 
For small perturbations, laminar motion is observed as far as $Re
\approx 10^{5}$ but the transition in pure fluid can be reached for
$Re \approx 2100$ provided that the perturbation is strong enough to
allow the growth of turbulent ``puffs'' \cite{Tritton}.  Recent
studies have investigated with different kinds of perturbations the
nature of the unstable modes, either in the inlet region or in the
fully developed flow \cite{Eliahou, Darbyshire}.

The objective of the present work is to examine how transition to
turbulence is affected by the presence of suspended particles in the
simplest case of neutral buoyancy.  More specifically, we focus upon
determining the transition threshold between the laminar and the
intermittent regime as a function of the particle volume fraction
$\phi$ of the suspension.  Because the particles are neutrally buoyant
and largely drag-free, the present study is related to recent work
which has examined global subcritical stability behavior of plane
Couette flow forced by the presence of a single spherical bead or a
spanwise wire \cite{Bottin, Bottin2}.  This work also has a practical
aspect as it is related to pipeline flow of slurries.

Experiments are performed with four sets of spherical polystyrene
particles having density $\rho~=~1.0510~\pm~0.001$~g.cm$^{-3}$ and
diameters $d$ presented in table \ref{tab:particles}.  To obtain
neutral buoyancy, the densities of the fluid and of the particles are
matched.  We choose as a fluid a mixture of $22~\%$ glycerol and
$78~\%$ water by mass.  The temperature of the mixture is maintained
at $25~\pm~1^{\circ}~\mathrm{C}$ by using a thermostated bath as a
fluid reservoir in the fluid circulating loop.  At this temperature,
the viscosity of the mixture is $\mu~=~1.64~\pm~0.03$~cP.

\begin{table}
    \begin{ruledtabular}
    \begin{tabular}{ccccc}
        \hline
        $d$ ($\mu$m) & $40~\pm~6$ & $215~\pm~25$ & $510~\pm~60$ &
        $780~\pm~110$ \\
        $D_{1}/d$ & $200~\pm~26$ & $37~\pm~4$ & $16~\pm~2$ & 
$10~\pm~1$ \\ 
        $D_{2}/d$ & $350~\pm~50$ & $65~\pm~7$ & $27~\pm~3$ & 
$18~\pm~2$ \\ 
    \end{tabular}
    \end{ruledtabular}
\caption{Particle diameters and pipe to particle diameter ratios.  The
smaller particles were supplied by Kodak (Rochester, NY USA) while the
others were from Maxiblast (South Bend, IN USA).}
\label{tab:particles}
\end{table}

The experimental set-up consists of a straight and horizontal
cylindrical glass tube of $2.6$~m length mounted on a rigid support
structure.  Two different tubes having different inner diameters
$D_{1}=8$~mm and $D_{2}=14$~mm are used in the experiments.  These
tubes are longer than the entry lengths necessary for the laminar flow
to fully develop at $Re \approx 2000$, $L_{e}(D_{1}) \sim 0.6$~m and
$L_{e}(D_{2}) \sim 1$~m.  The pipe to particle diameter ratios,
$D/d$, used in the experiments by combining the different particles
and tubes are indicated in table \ref{tab:particles}.  In order to
ensure that the flow in the pipe is undisturbed by perturbations from
a pump, the flow is driven by gravity.  The suspension is delivered to
the tube by overflow from a tank positioned at a fixed height to an
outlet of variable height, passing through the glass tube.
A Moineau progressing cavity pump (PCM model MR2.6H24) carries the
suspension back to the overflowing tank, but is isolated from the 
flow 
through the glass tube.

With these flow conditions, the transition for pure fluid was found to
take place at $Re \approx 2300$, but the flow was very sensitive to
any kind of perturbation applied to the pipe.  In order to control the
transition with a known perturbation, a ring (solid annulus) fitting
tightly into the tube and of thickness $1.5$~mm was inserted in the
pipe at its entrance.  The perturbation produced by the ring then
controlled the transition and lowered it down to $Re \approx 2100$, 
which is
the lowest value that could be reached with this kind of 
perturbation. 
The influence of the particles on the transition has been studied with
this ring.

The different flow regimes, i.e. laminar, intermittent or
turbulent, are identified by visualization as well as by measuring
the pressure drop between the entrance and the exit of the glass tube.
 
A classical way to detect the pipe-flow transition for pure fluid is 
to
inject dye and to observe the wavering of the streak as it passes down
the pipe \cite{Reynolds, Tritton}.  In the present experiment with
particles, direct visualization of the particle motion in the
suspension is a very straightforward way of observing the birth and
convection of ``turbulent puffs'' in the flow.  Whereas motion in the
laminar regime is characterized by parallel particle trajectories,
motion in a turbulent puff presents a strong mixing in the radial
direction.  When the flow rate is increased, we observe more and more
turbulent puffs in the flow, and finally these merge once
the turbulent regime is established.
  	
\begin{figure}[h]
  \centering
  \includegraphics[scale=0.45]{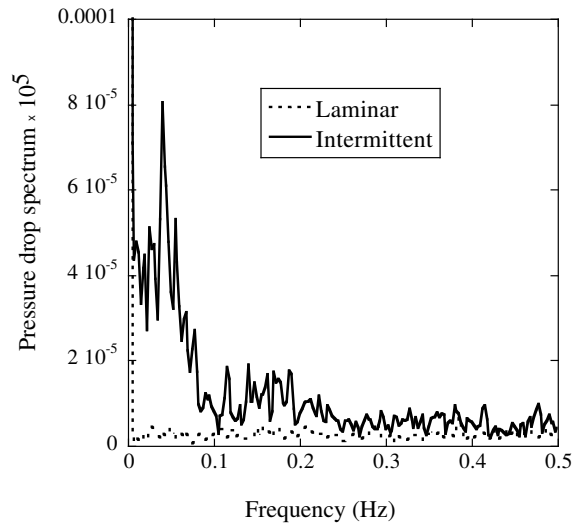} 
  \caption{Fluctuations of the pressure drop across the pipe for the
  laminar and the intermittent regime in a suspension of particle
  volume fraction $\phi= 0.1$.  The zero frequency value of the
  spectrum, related to the mean value of the pressure drop, is much
  larger and not represented on the graph}
  \label{fig:spectra}
\end{figure}

To provide a more quantitative indicator of transition, we also
measured the pressure drop between the entrance and the exit of the
glass tube with electronic manometers (Newport Omega PX 154). 
Pressure fluctuations due to the larger pressure drop caused by the
turbulent puffs could then be clearly identified with a signal
analyzer (Hewlett Packard 3562A).  Figure \ref{fig:spectra} shows
pressure drop spectra in the low-frequency range for the laminar flow
and at the onset of intermittency, both for a particle volume fraction
$\phi = 0.1$.  The spectrum in the intermittent regime displays strong
fluctuations for frequencies under $0.1$~Hz.  Using the spectral
signature, the onset of intermittency can thus be clearly identified
even for large volume fractions where the visualization is not
possible.  The spectrum of pressure fluctuations in the flow driven by
the pump was observed to consist of ``continuous noise'' of the type
shown by figure \ref{fig:spectra}, plus a strong peak at the frequency
of the pump rotation with its associated harmonics.  By contrast, the
fluctuations when the flow is driven by gravity are observed to have
only the ``continuous noise''.  Furthermore, this noise was determined
to be unchanged when the pump was off and the overflow tank was filled
manually.  Consequently, the spectrum observed in the gravity-driven
flow is considered to be independent of the pump.

When the threshold of intermittency was reached, we measured the
critical flow rate $Q_{c}$ and deduced the critical Reynolds number
$Re_{c} = 4 Q_{c}\rho /\pi \mu D$ with the pipe diameter $D$ as the
length scale.  To measure $Q$, we collected a given volume of the
suspension without altering the flow, by capturing fluid at the outlet
to the thermostatically controlled reservoir.  The time needed to
obtain this volume yielded $Q_{c}$.  The particles collected in this
volume were then sieved, rinsed, dried, and weighed to provide the
mean particle volume fraction $\phi$ in the suspension flow. 
Measurements of the critical Reynolds number associated with the start
of intermittency have been carried out for different combinations of
particles and tubes presented in table \ref{tab:particles} and for
different concentrations $\phi$ ranging from $0$ to $0.3$.
  
\begin{figure}[h]
  \centering
  \includegraphics[scale=0.48]{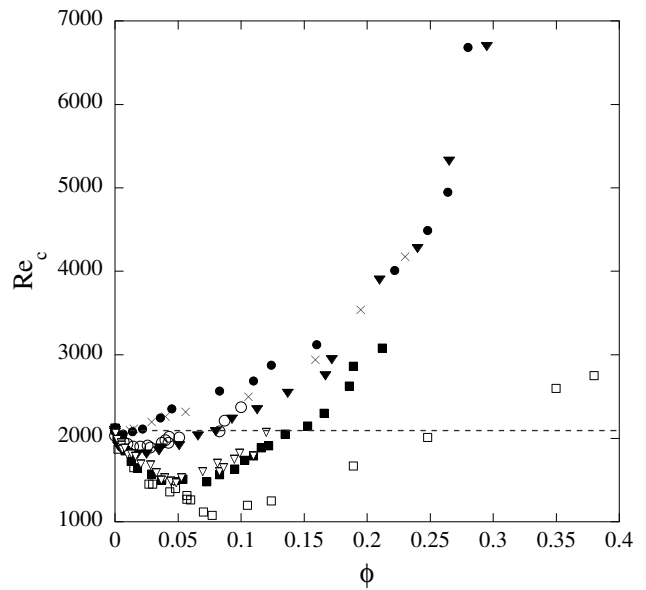}
  \caption{Critical Reynolds number $Re_{c}$ as a function of the
  volume fraction $\phi$ of the suspension for different combinations
  of particles and tubes.  Tube $D_{1}$ with $d~=~215~\mu$m ($\circ$),
  $510~\mu$m ($\triangledown$), $780\mu$m ($\square$).  Tube $D_{2}$
  with $d~=~40~\mu$m ($\times$), $215~\mu$m ($\bullet$), $510~\mu$m
  ($\blacktriangledown$), $780~\mu$m ($\blacksquare$).}
  \label{fig:results}
\end{figure}

Figure \ref{fig:results} displays the critical Reynolds number
$Re_{c}$ as a function of the particle volume fraction $\phi$ for the
different combinations of particles and tubes.  The results indicate
two different situations:

\begin{itemize}
    
\item[(i)] For particles with $D/d \geq 65$ ($\times$ and $\bullet$),
the transition is shifted to larger Reynolds numbers.  The critical
Reynolds number $Re_{c}$ is a monotonically increasing function of
$\phi$.  Furthermore, the data for the two sizes collapse onto a
single curve, so that at a given $\phi$ in that range of $D/d$, the
transition threshold does not depend on the diameters of the particles
and of the tube.
      
\item[(ii)] For particles with $D/d \leq 65$, the behavior depends on
$\phi$.  For small $\phi$, the transition is moved to lower Reynolds
number and $Re_{c}$ decreases with increasing $\phi$.  The magnitude
of the decrease and in particular the minimum $Re_{c}$ reached as
$\phi$ is increased depends on the diameters of the particles and the
tube.  In particular, for the 780 $\mu$m particles in pipe $D_{1}$ for
which $D/d \sim 10$ ($\square$), intermittency is observed at $Re_c
\sim 1000$ for $\phi \sim 0.07$.  This is the smallest $Re_c$ found in
this work.  For larger $\phi$, $Re_{c}$ increases with increasing
$\phi$ and the transition is eventually delayed.  It is worth noticing
that the data for combinations of particles and tube having roughly
the same $D/d$ collapse onto the same curve ($\circ$ and
$\blacktriangledown$ for $D/d \sim 27-37$ and $\triangledown$ and
$\blacksquare$ for $D/d \sim 16-18$). 

\end{itemize}

The behavior for $D/d \geq 65$ seems simpler since it is independent
of the diameters of the particles and of the tube.  The observed delay
to the transition is expected to be due to an enhancement of
the viscosity caused by the suspended particles.  Common models for
the \emph{effective} viscosity $\mu_{e}$ of a suspension do not depend
on the size of the particles and give an increase of the viscosity
with increasing average concentration $\phi$.  A well-known example is
Krieger's viscosity \cite{Krieger,Phillips}, which expresses the
effective viscosity $\mu_{e}$ as a function of $\phi$ according to the
law:
\begin{equation}
   \frac{\mu_{e}}{\mu}= \left(1-\phi/\phi_{m}\right)^{-1.82}
   \label{eq:krieger}
\end{equation}
where $\mu$ is the viscosity of the pure fluid and $\phi_{m} = 0.68$
the random close packing (i.e.~maximum) concentration for spherical
particles.  This empirical formula initially determined at low
Reynolds number is commonly used at finite Reynolds.

\begin{figure}[h]
  \centering
  \includegraphics[scale=0.48]{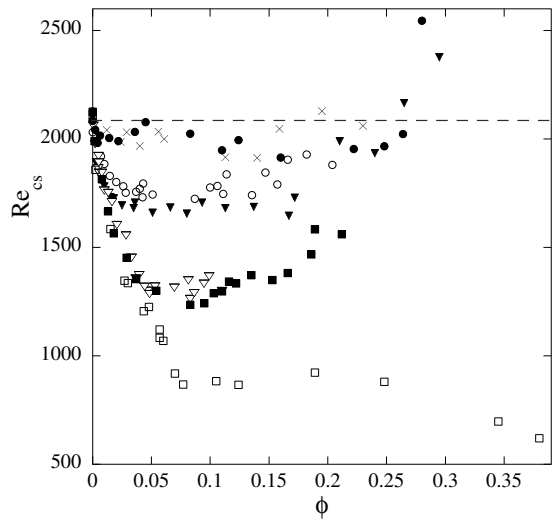}
  \caption{Critical Reynolds number of the suspension using Krieger's
  viscosity $Re_{cs}=Re_{c} \mu_{0}/\mu_{e}$ as a function of particle
  volume fraction $\phi$.  Tube $D_{1}$ with $d~=~215~\mu$m ($\circ$),
  $510~\mu$m ($\triangledown$), $780\mu$m ($\square$).  Tube $D_{2}$
  with $d~=~40~\mu$m ($\times$), $215~\mu$m ($\bullet$), $510~\mu$m
  ($\blacktriangledown$), $780~\mu$m ($\blacksquare$).}
  \label{fig:krieger}
\end{figure}

The critical Reynolds number of the suspension using equation
(\ref{eq:krieger}), $Re_{cs}=Re_{c} \mu/\mu_{e}$, is plotted as a
function of $\phi$ in figure \ref{fig:krieger}.  We observe that for
$D/d \geq 65$ ($\times$ and $\bullet$) $Re_{cs}$ is approximately
independent of $\phi$ and remains close to the value for the pure
fluid, $Re_{cs} \approx 2100$. For $\phi \geq 0.25$, we observe a 
steep 
increase in this threshold ($\bullet$ and also 
$\blacktriangledown$), suggesting 
the presence of an additional mechanism 
for dissipation beyond the viscosity 
enhancement observed in Stokes 
flow suspensions. 
However, we have only a limited range of concentration data, because
the large flow rates required to achieve transition at the elevated
effective viscosity result in a large pressure drop.  The large
pressure exceeds the range of the pressure gauges and also
necessitates the direct use of the pump instead of gravity-driven 
flow 
that we have chosen to consider only in this study. 

\begin{figure}[h]
  \centering
  \includegraphics[scale=0.48]{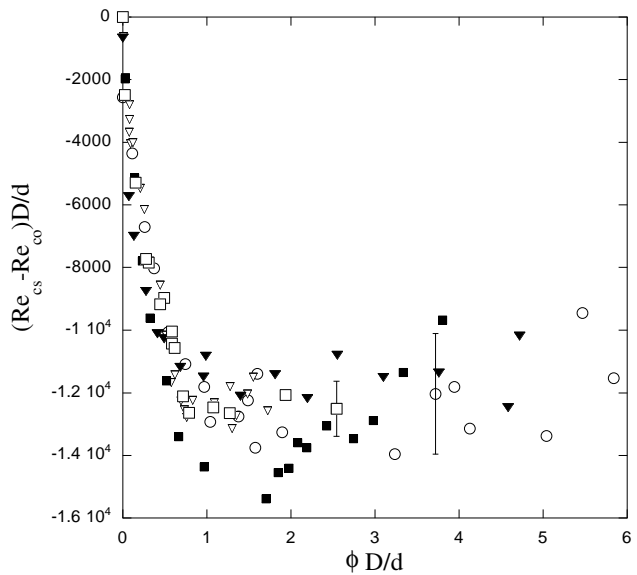} 
\caption{Scaled difference between the critical Reynolds number of the
suspension using Krieger's viscosity $Re_{cs}$ and that of the pure
fluid $Re_{c0}=2100$, i.e. $(Re_{cs}-Re_{c0}) D/d$ as a function of
$\phi D/d$ for $\phi<0.25$.  The values of the parameters are as in figure
\protect\ref{fig:krieger} (only the data for particles which cause 
earlier transition have been kept).}
  \label{fig:scaling}
\end{figure}

This simple scaling with Krieger's viscosity is not sufficient to
obtain a collapse of the curves for $D/d \leq 65$ (see figure
\ref{fig:krieger} for $\circ$, $\triangledown$, $\square$,
$\blacktriangledown$, $\blacksquare$).  However, we can notice a
similarity between these curves.  At low $\phi$, there is a first
regime where $Re_{cs}$ decreases steeply with increasing $\phi$ and
the different curves appear to collapse.  It should be mentioned that
a very small decrease is also observed for the 215~$\mu$m particles in
pipe $D_{2}$ ($\bullet$) but it can be considered to be within error
bars. Above a critical volume fraction $\phi_{c}$ which depends on 
the pipe 
to particle diameter ratio $D/d$, there is a second regime 
where the
curves eventually reach minimum values of $Re_{cs}$ and remain
approximately independent of $\phi$ for larger volume fractions.  
These
minimum values of $Re_{cs}$ decrease with increasing $D/d$.  Moreover,
the data for combinations of particles and tube having similar values
of $D/d$ seem to collapse onto the same curve.  These last
observations lead us to scale the difference between the critical
Reynolds number of the suspension $Re_{cs}$ and that of the pure fluid
$Re_{c0}=2100$ as well as $\phi$ with $D/d$ as displayed in figure
\ref{fig:scaling}.  This new scaling provides a collapse of the
different curves for all $D/d$.  The scaled difference
$(Re_{cs}-Re_{c0}) D/d$ initially decreases rapidly and approximately
linearly with $\phi D/d$, then saturates and eventually increases
slightly at larger $\phi D/d$. The results on figure
\ref{fig:scaling} suggest $\phi_{c} \sim d/D$.

These data were obtained with the ring at the entrance of the pipe but
similar results were also observed without it.  This suggests that the
subcritical transition is triggered by the particles.  A particle
introduces fluctuational velocities whose form and coupling to the
mean flow vary with the particle Reynolds number $Re_{p}=Re \, d^{2}/
D^{2}$.  Unlike the case of uniform flow, the influence of $Re_p$ on
the structure of the dual wakes caused by shear flow around suspended
bodies is, while under study \cite{Zettner, Bottin}, not well
understood.  It is plausible that the influence of particle size in
reducing $Re_{cs}$ at a fixed $\phi$ results from the increase in the
particle scale Reynolds number with $d/D$.  With increasing $Re_p$,
the disturbance flow caused by the particle is presumably less
efficiently dissipated by viscous action, thus allowing for stronger
coupling to the bulk flow.

A more refined analysis should take into account the influence of
particles on the velocity and concentration profile in the pipe flow. 
In conditions of very low Reynolds number and high concentration,
particles migrate towards the center of the pipe and blunt the
velocity profile, see for instance \cite{Hampton}.  This effect which 
would tend to reduce the effective viscosity in the pipe is suspected 
to be
present for the 780~$\mu$m particles in pipe $D_{1}$ for $\phi \geq 
0.25$ (see $\square$ in figure \ref{fig:krieger}).  There exists a 
second type of migration, the so-called
``tubular pinch effect'', which is inertial and causes a single
particle to move to a position at a distance of $0.3 D$ from the axis
\cite{Segre, Han}.  This effect is observed with the larger particles
in both pipes.  The present study seems to suggest that the particles
alter the threshold of the subcritical transition through coupling of
the base flow to velocity fluctuations rather than the base flow
itself through their migration but this requires confirmation.

This work leads to definite conclusions regarding the influence of
suspended neutrally-buoyant solids upon the transition away from
laminar flow.  The influence depends both upon the pipe to particle
diameter ratios and concentration.  For $D/d \geq 65$,
neutrally-buoyant particles cause a delay in transition to larger
$Re_c$.  This effect can be explained by the enhancement of the
effective viscosity of the suspension.  However, for $\phi>0.25$, the
delay in transition is found to be substantially larger than can be
explained by a simple renormalization by effective viscosities.  For
$D/d \leq 65$, the behavior is quite different and neutrally-buoyant
particles alter the transition to turbulence in pipe flow to smaller
$Re_c$.  Scaling the departure from the critical Reynolds number for
pure fluid as well as the concentration with $D/d$ gives a master
curve for the transition for all $D/d$.  The most plausible
explanation for the reduction of the critical Reynolds number, though
one still in need of confirmation, is that the fluctuations induced by
the particles trigger the subcritical transition.

\begin{acknowledgments}
We wish to thank P. Manneville for discussions and F. Ratouchniak for
technical assistance.  The donation of the 40$\mu$m particles secured
by Derin Adebekun of Kodak is greatly appreciated.  We would like also
to acknowledge support for this work from the Institut Fran\c{c}ais du
P\'etrole.  This study was also undertaken under the auspices of a
CNRS-NSF collaborative research ``Flow, resuspension, and
sedimentation of a suspension in a tube''.  Fellowship from
the French Minist\`ere de la Recherche is gratefully acknowledged by
J.-P. Matas.
\end{acknowledgments}

\end{document}